# Introducing an Instructional Model for Teaching Blended Math-Science Sensemaking in Undergraduate STEM Courses Using Computer Simulations

## Introduction

The ability to express scientific concepts in mathematical terms, and more generally integrate scientific and mathematical reasoning about a phenomenon, is a foundational cognitive process that is at the heart of scientific thinking (Redish, 2017; Zhao & Schuchardt, 2021; Kuo et al., 2013). This cognitive process called "blended math-science sensemaking" is a necessary skill for scientists, but it is equally necessary for all citizens to be able to appropriately use the results and reasoning of science to make better decisions in their jobs and in their personal lives. Only through this blending does science achieve its vital precision and predictive power. While the value of such sensemaking is well recognized, it is also clear that it is rare in the population. Few students are learning it from their STEM courses (Becker & Towns, 2012; Bing & Redish, 2009; Taasoobshirazi & Glynn, 2009; Tuminaro & Redish, 2007), and there is little research on how to teach it effectively.

In this work we introduce the development and testing of a novel instructional method for teaching blended math-science sensemaking (MSS) that is suitable for use in STEM courses in undergraduate and K-12 educational settings. This study builds on our past work on developing and validating a framework for characterizing in detail the cognitive levels involved in such sensemaking (Kaldaras & Wieman, 2023). It also employs the insights we gained as to the unique power of interactive simulations for assessing and developing blended MSS.

Sensemaking is a *dynamic process* of building or revising an explanation to ascertain the mechanism underlying a phenomenon in order to resolve a gap or inconsistency in one's understanding (Oden & Russ, 2017). Blended MSS is reflected in deep conceptual understanding of quantitative relationships and scientific meaning of equations (Zhao & Schuchard, 2021; Kuo et al., 2013). This type of sensemaking has been shown to be a critical component of expert-like understanding and expert mental models in any scientific discipline (Zhao & Schuchardt, 2021) as well as a critical component required for developing authentic problem-solving skills (Schoenfeld, 2016; Wieman et al., 2008). Various aspects of the blended MSS have been described for specific disciplines (Bing & Redish, 2007; Tuminaro & Redish, 2007; Ralph & Lewis, 2018; Lythcott, 1990; Schuchardt & Schunn, 2016; Hunter et al., 2021). Further, different ways of engaging in blended MSS have been described in literature focusing on a cognitive framework that outlines the process of blended MSS and relating it to basic cognition (Gifford & Finkelstein, 2020).

Despite a significant body of research on the subject, blended MSS is not well understood at the practical level, and thus not explicitly taught as a result. However educational support is needed to help students develop blended MSS skills, which in turn is necessary for developing a deep understanding of science. Prior work has been conducted on creating instructional approaches that support authentic learning of physics focused on investigative exploration of natural phenomena through quantitative observational experiments (Etkina et al., 2021, 2020; Karelina et al., 2007). This approach, however, does not explicitly support students in developing an understanding of why specific mathematical relationships are appropriate for describing a given natural phenomenon and relating all the aspects of the relationship to specific observations of the natural phenomenon described by it. This ability to provide a causal explanation for the equation structure in relation to the natural phenomenon it describes lies in the center of the blended MSS process (Kaldaras & Wieman, 2023) and is the focus of the instructional approach described in this study. Moreover, the instructional approach introduced in this study is grounded in previously validated cognitive framework for blended MSS (Kaldaras & Wieman, 2023), which provides a roadmap for developing learning sequences focused on engaging students in blended MSS at the increasing levels of sophistication and across various STEM disciplines. Grounding the learning sequence in a validated cognitive framework ensures that the learning sequence targets specific research-based skills and knowledge needed to develop higher proficiency in blended MSS. Furthermore, aligning learning sequences to specific levels of the validated cognitive framework allows for quick and efficient diagnosis of the types of blended MSS students are struggling with in a given context and provides information on the supports students need to overcome these challenges and develop blended MSS proficiency. The current study provides an instructional model for supporting blended MSS across various STEM disciplines, particularly Physics and Chemistry

We believe that including supportive activities that target blended MSS in science courses will benefit all students and could remove barriers for underrepresented students: research has shown that such barriers might be tied to insufficient preparation in Math (Foltz et al., 2014; Green et al., 2018; Ralph & Lewis, 2018). In this study we are introducing an instructional model for designing instructional materials and assessments that don't just support the development of routine Math application skill but foster the ability to interpret math in science for explaining phenomena and solving real-life problems.

Designing effective ways of teaching and assessing any construct requires understanding of how proficiency in that construct develops with time (National Research Council [NRC], 2001). A construct refers to a coherent collection of concepts and skills that can be used to explain performance on a test for a given topic. Proficiency in a construct refers to describing what mastery in that construct looks like (NRC, 2001). The understanding of how proficiency in a construct develops is essential for designing effective instructional and assessment strategies, empirical testing and valid interpretation of assessment results and aligning curriculum, instruction and assessment with the purpose of helping students achieve higher proficiency in a construct (NRC, 2001).

Generally, cognitive models describe how students represent knowledge and develop proficiency in a construct (NRC, 2001). Little work has been done on formulating and testing a theory of blended MSS that (1) outlines proficiency levels of this cognitive concept and (2) applies across different scientific fields. Initial work was done by Zhao and Schuchardt (2021) that proposed a theoretical cognitive model for independent mathematics and science dimensions outlining increasingly sophisticated proficiency levels. Their model is grounded in a review of relevant literature across different fields (including math, physics, chemistry, biology) and represents mathematical and scientific sensemaking as separate dimensions. Building off this work, we developed and validated a cognitive framework for blended MSS proficiency levels (Kaldaras & Wieman, 2023). We have found that student blended MSS proficiency is largely independent of the specific disciplinary context (Kaldaras & Wieman, 2023).

In this study we are using our framework for blended MSS as a guide to develop instructional sequences in Physics and Chemistry to support students in developing blended MSS cognitive skills. The instructional sequences we developed are based on PhET simulations (sims) (PhET Interactive Simulations). PhET sims represent a uniquely suitable tool for helping build instructional sequences that provide an authentic learning environment for fostering blended MSS. Specifically, sensemaking is a dynamic process (Odden & Russ, 2017) that focuses on the interplay of action and interpretation of the results of the action (Weick et al., 2005). Therefore, supporting the process of sensemaking, including blended MSS, calls for a dynamic and interactive learning environment that allows for continuous accumulation of new quantitative evidence and feedback associated with changing the parameters of the system in question in order to support revising the explanations and developing quantitative accounts of the phenomena. PhET sims possess all these features and offer great potential for creating effective learning environments for supporting blended MSS. PhET sims can be enhanced by coupling them with relevant instructional materials that include additional information (such as data, reading materials etc.). PhET sims offer a wide range of disciplinary contexts. Over the years, the PhET sims project has developed numerous simulations across various fields of science; they are free and globally accessible and used nearly 1 million times/day. All these features make PhET sims a uniquely useful tool for designing instructional sequences that can effectively support blended MSS.

We believe that the instructional model introduced in this paper will help develop blended MSS proficiency in the classroom. This work presents the first attempt to develop research-based instructional materials to support proficiency development of blended MSS grounded in the validated cognitive framework. The two research questions (RQ) addressed in this paper are:

*RQ 1: Are the instructional sequences developed following the instructional model grounded in the cognitive framework introduced in this study effective in helping students develop higher proficiency in blended MSS?*

*RQ2: What instructional elements are helpful in supporting students' progress along the levels of the cognitive model?*

We will discuss the implementation and results of using these sequences in introductory undergraduate chemistry and physics classrooms. We will focus specifically on discussing student learning outcomes in terms of student progress along the levels of the cognitive framework, which will help answer RQ 1. We will also discuss specific instructional scaffolds and tasks that supported students' progress towards higher levels of the cognitive framework, which will answer RQ 2.

**Theoretical Framework**

*Elements of an Effective Learning System*

Cognitive Framework for Blended Math-Sci Sensemaking

Effective learning systems support development of higher proficiency in a topic by offering cognitively appropriate learning supports. Prior research suggests using validated cognitive frameworks as roadmaps for building such learning systems (NRC, 2001, 2012; 2014; Alonso & Gotwals, 2012; Kaldaras et al., 2021). Cognitive frameworks describe how proficiency in a domain develops by outlining increasingly sophisticated ways of thinking about a construct (NRC, 2000). Aligning learning systems to validated cognitive frameworks ensures that students are provided with research-based scaffolding aimed at helping them build deeper understanding of a construct coherently over time as they progress through the learning system (Alonso & Gotwals, 2012; Kaldaras, 2020).

The learning sequences developed in this study are aligned with our previously validated cognitive framework for blended Math-Science sensemaking (Kaldaras & Wieman, 2023) that reflects increasingly sophisticated ways of blending Math and Science cognitive dimensions to make sense of scientific phenomena mathematically. The cognitive framework is shown in Table 1. The framework describes increasingly sophisticated ways of engaging in blended Math-Science sensemaking as students are working towards developing a mathematical formula describing the scientific phenomenon in question or building a deeper understanding of the known formula. The model consists of three broad levels: qualitative (level 1), quantitative (level 2) and conceptual (level 3). Each broad level consists of three sub-levels: "Description", "Pattern", and "Mechanism". At the qualitative level students can't develop the exact mathematical relationship describing the scientific phenomenon in question, but they can identify the relevant variables ("Description"), qualitative patterns among the variables ("Pattern") and describe qualitative causal mechanism of the phenomenon ("Mechanism"). At the quantitative level students can identify numerical values of the relevant variables ("Description"), quantitative patterns among the variables ("Pattern") and develop a mathematical relationship describing the phenomenon and justify the equation using numerical values of the variables ("Mechanism"). At the conceptual level students justify the scientific need to include all unobservable variables and constants into the mathematical relationship ("Description"), justify the mathematical relationship by relating the observed quantitative patterns to specific mathematical operations ("Pattern") and describe the causal mechanism of the phenomenon reflected in the equation structure ("Mechanism").

The cognitive framework shown in Table 1 was used as a guide in the current study to develop the instructional sequences. All the tasks in the learning sequences were designed to align to specific levels of the framework shown in Table 1. In the two learning sequences designed for this study students were working towards developing a mathematical relationship for a scientific phenomenon in question. The tasks were designed to engage students in blended MSS at the specific sub-levels of the framework starting from the lowest and progressing to the highest level at the end of the activity.

<u>Using Computer Simulations to Support Engagement in Blended Math-Sci Sensemaking</u>

Effective engagement in blended MSS calls for interactive systems that enable dynamic acquisition and analysis of quantitative data that support building a deeper understanding of quantitative relationships describing phenomena. PhET sims represent a suitable system for designing such learning environments. The design features embedded in PhET sims allow to change the numerical values of the parameters and observe how the imposed changes affect the numerical values of other parameters and the phenomenon in the simulation. PhET sims are easily supplemented with numerical data if needed either by collecting the data from the simulation directly, or by using other numerical data that aligns with the sim The instructional approach presented here leveraged the capabilities of PhET sims to create learning environments that supports authentic engagement in blended MSS process at various levels of sophistication as reflected in the cognitive framework shown in Table 1.

*Table 1. Theoretical Blended Math-Sci Sensemaking Framework[1]*

| 1 Qualitative | Description | Students can use observations to identify which measurable quantities (variables) contribute to the phenomenon.<br>*Example: force and mass make a difference in the speed of a car.* |
|---|---|---|
| | Pattern | Students recognize patterns among the variables identified using observations and can explain *qualitatively* how the change in one variable affects other variables, and how these changes relate to the scientific phenomenon in question.<br>*Example: the smaller car speeds up more than the big car when the same force is exerted on both.* |
| | Mechanism | Students demonstrate *qualitative* understanding of the underlying causal scientific mechanism (cause-effect relationships) behind the phenomenon based on the observations but can't define the exact mathematical relationship.<br>*Example: it is easier to move lighter objects than heavy objects, so exerting the same force on a lighter car as on a heavy car will cause the lighter car to speed up faster.* |
| 2 Quantitativ | Description | Students recognize that the variables identified using the observations provide measures of scientific characteristics and can explain *quantitatively* how the change in one variable affects other variables (but not recognizing the quantitative patterns yet), and how this change relates to the phenomenon. Students not yet able to express the phenomenon as an equation.<br>*Example: recognizing that as variable A changes by 1-unit, variable B changes by 2 units.* |
| | Pattern | Students *recognize quantitative patterns* among variables and explain *quantitatively (in terms of an equation or formula)* how the change in one parameter affects other parameters, and how these changes relate to the phenomenon in question. Students not yet able to relate the observed patterns to the operations in a mathematical equation and can't develop the exact mathematical relationship yet.<br>*Example: recognizing mathematical relationships such is direct linear and inverse linear among others* |
| | Mechanism | Students can explain *quantitatively* (express relationship as an equation) for how the change in one variable |

[1] Note that the examples provided in the table assume students are working towards developing a mathematical relationship describing the scientific phenomenon in question.

| e | | affects other variables based on the quantitative patterns derived from observations. Students include the relevant variables that are not obvious or directly observable. Students not yet able to explain conceptually why each variable should be in the equation beyond noting that the specific numerical values of variables and observed quantities match with this equation. Students cannot explain how the mathematical operations used in the equation relate to the phenomenon, and why a certain mathematical operation was used. Students can provide causal account for the phenomenon. <br> _Example_: _In $F_{net}=ma$, multiplication makes sense because when applied force on the mass of 50 kg increases from 10 to 20 N, acceleration increases by 2. That only makes sense for a multiplication operation._ |
|---|---|---|
| **3** <br> C o n c e p t u a l | Description | Students can describe the observed phenomenon in terms of an equation, and they can explain why all variables or constants (including unobservable or not directly obvious ones) should be included in the equation. Students not yet able to explain how the mathematical operations used in the formula relate to the phenomenon. <br> _Example_: _In $F=ma$, the F is always less than applied force by specific number, so there must be another variable subtracted from $F_{applied}$ to make the equation work. The variable involves the properties of the surface. So, the equation should be modified: $F_{applied}-(variable)=ma$_ |
| | Pattern | Students can describe the observed phenomenon in terms of an equation, and they can explain why all variables or constants (including unobservable or not directly obvious ones) should be included in the equation. Students not yet able to provide a causal explanation of the equation structure. <br> _Example_: _In $F_{net}=ma$, multiplication makes sense because as applied force on the same mass increases, acceleration increases linearly, which suggests multiplication._ |
| | Mechanism | Students can describe the observed phenomenon in terms of an equation, and they can explain why all variables or constants (including unobservable or not directly obvious ones) should be included in the equation. Students can fully explain how the mathematical operations used in the equation relate to the phenomenon in questions and therefore provide causal explanation of the equation structure, that is how the equation (the variables and the mathematical operations among the variables) is describing the causal mechanism of the scientific phenomenon. <br> _Example_: _since greater acceleration is caused by applying a larger net force to a given mass, this shows a positive linear relationship between a and $F_{net}$, which implies multiplication between m and a in the equation, or $F_{net}=ma$._ |

**Methods**

*Choosing Science Topics*

We chose the topics of Coulomb's law for Physics and Heat Capacity for Chemistry classes after discussion with the respective course instructors. These topics were chosen because they were part of the corresponding course curriculum, allowed for meaningful engagement in blended MSS at all the levels of the cognitive framework shown in Table 1, and had an associated PhET simulation available that could be used in the learning activity. The screenshots of the simulations are shown in Figures 1 and 2 respectively. The Coulomb's law simulation allows students to explore the relationship between the amount and the type of charge, the distance between charges and the associated attractive or repulsive electric force. The target formula for the Coulomb's Law activity was [$F = (Q_1 \times Q_2/distance^2) \times$ Coulomb's constant], where "F" is electric force exerted by one charge on the other, "$Q_1$" and "$Q_2$" are the magnitudes of each charge, "distance" is the distance between charges and "Coulomb's constant" is a proportionality constant specific to the medium in which the two charges are interacting. The Energy Forms and Changes simulation allows one to explore the relationship between the amount of energy needed to raise the temperature of various substances. The target formula for the Heat Capacity activity was [ Energy required to change the T of a given amount of substance = C × **Δ**T× m], where "C" is specific heat capacity of a substance, "**Δ**T" is the change in temperature, and "m" is mass of the substance.

*Figure 1. Screenshot of Coulomb's Law simulation*

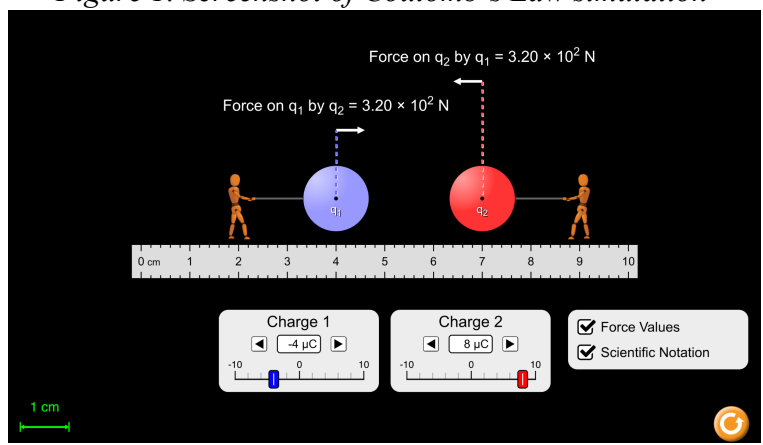

*Figure 2. Screenshot of Energy Forms and Changes Simulation*

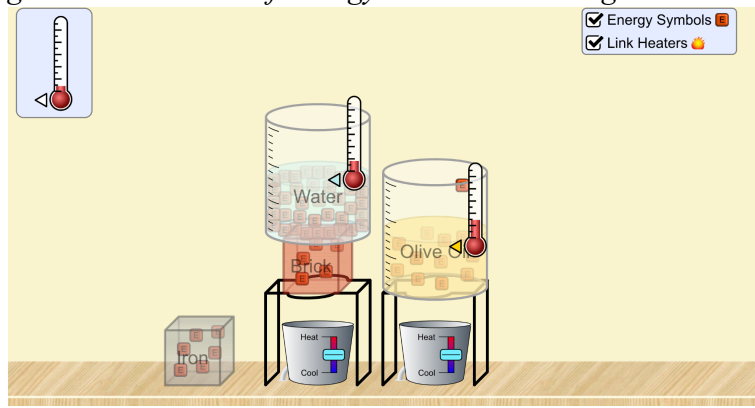

*Learning Activity Design Principles*

The goal of both learning activities was to have students develop a mathematical relationship for the topic of interest: Coulomb's law for Physics and Heat Capacity for Chemistry. Table 2 shows the general activity structure that we followed in designing both activities. The full activities for Physics and Chemistry are available in the Appendix. This structure and the design principles discussed below are widely applicable for developing learning activities supporting blended MSS across STEM disciplines at the K-12 and undergraduate levels.

*Table 2. Learning Activity Structure for Supporting Blended Math-Sci Sensemaking*

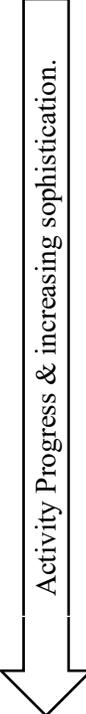

| Activity Task[1] (Brief description) | Cognitive Framework Alignment |
|---|---|
| Pre-Assessment (CR[2]) | Assess the starting level before the activity |
| Identify the variables relevant for characterizing the phenomenon mathematically (MC[3]) | Level 1 ("Description") |
| Identify qualitative patterns among the variables (MC) | Level 1 ("Pattern") |
| Identify quantitative patterns among the variables using the simulation and the data when applicable (MC) | Level 2 ("Pattern") |
| Mid-Assessment: suggest and justify a mathematical relationship for the observed phenomenon using what you have learned so far in the activity (CR) | Assess the level upon exploring quantitative patterns |
| Post-Assessment: pick the most likely mathematical relationship for the observed phenomenon (MC) and justify your choice using what you have learned in this activity (CR) | Assess the level upon exploring quantitative patterns |
| Suggest how to calculate the value of the relevant constants (if applicable) and explain the scientific meaning of the constants (CR)[4] | Level 3 ("Description") |
| Describe the causal mechanism for the phenomenon reflected in the mathematical relationship that you proposed | Level 3 ("Mechanism") |
| Application questions for a given topic (if needed) | Assess whether level placement is consistent across contexts (transfer) |

(Left margin, vertical text): Activity Progress & increasing sophistication.

[1] Students are engaging with each task as they are interacting with the simulation.
[2] Constructed Response
[3] Multiple Choice
[4] Coulomb's law constant and Specific Heat constant for Physics and Chemistry activities respectively

The design of the learning activities incorporated the elements of an effective learning system discussed in the Theoretical Framework section. Specifically, we used the cognitive framework for blended MSS (Table 1) as a guide to develop both learning sequences and leverage the design features of PhET sims in the activity design. As shown in Table 2, the activity tasks were aligned to the sub-levels of the cognitive framework. This ensured that the learning sequences followed a developmental approach and supported student engagement in blended MSS according to the increasing levels of sophistication in the cognitive framework. Additionally, we leveraged the PhET sims design features to structure student engagement in blended MSS at various levels of sophistication. For example, we asked students to use the simulation to identify the relevant variables (level 1 "Description"), qualitative and quantitative patterns among the variables (level 1 "Pattern" and level 2 "Pattern" respectively). Both simulations supported student engagement in blended MSS with these tasks by allowing students to change various conditions of the system and observe how the imposed changes affected the relevant variables. For example, in Coulomb's law sim the students were able to change the magnitude and the type of charges and the distance between them and observe how the magnitude of the associated attractive or repulsive electric force changes. Similarly, in Energy Forms and Changes sim the students were able to heat up various substances such as water, brick iron and oil and observe how much energy shown in the form of energy cubes it takes to raise the temperature of various substances by specific number of degrees.

Finally, both activities included three formative assessment tasks probing whether students have developed the mathematical relationship of interest at various points in the activity: 1) pre-assessment before engaging in the activity, 2) mid-activity assessment after exploring quantitative patterns using the simulation and the data (when applicable), 3) post activity assessment.

*Physics and Chemistry Activity Design Differences*

Both Chemistry and Physics activity followed a similar design structure discussed in the section above and shown in Table 2. However, there were several design features that were different between the two activities. These differences are summarized in Table 3 below.

*Table 3. Design differences between Physics and Chemistry activities.*

| Instructional design feature | Coulomb's law | Heat Capacity |
|---|---|---|
| Simulation design | Allows to explore specific quantitative relationships | Quantitative exploration is supported less. However, it is possible to count the number of energy cubes it takes to raise the temperature of a substance by 1 or more degrees. |
| Pre-assessment | Develop mathematical relationship based on introductory scenario before interacting with the simulation | Develop mathematical relationship based on interaction with the simulation |
| Data availability | Data was not provided since the simulation allowed for quantitative exploration | Data was provided since the simulation did not fully support quantitative exploration |
| Exploration of quantitative patterns and post-assessment | Asked to explore the quantitative patterns using the simulation prior to asking them to develop a mathematical relationship. | Asked to explore the quantitative patterns using the simulation prior to showing the data. Data was provided to students, but they were not explicitly asked to identify specific quantitative patterns in the data prior to asking them to develop a mathematical relationship |

The sims used in the two activities were different in the degree to which they allow for exploration of quantitative patterns. Coulomb's law sim allows for exploration of numerical quantitative patterns focused on how the magnitude of attractive and repulsive forces is affected by distance between the two changes and charge magnitude (see Figure 1). Energy forms and changes sim, in turn, is more qualitative, but it does allow to observe the number of energy units (energy cubes) it takes to heat up a given substance by a certain number of degrees (Figure 2). Because of these differences in the two sims, we provided data to students for Energy Forms and Changes sim (see Heat Capacity Activity in the Appendix) to ensure that students have the instructional supports for engaging in quantitative pattern identification for this phenomenon. For Coulomb's law activity students were not provided any data, but rather were expected to use the simulation to engage in quantitative pattern identification.

Further, in Coulomb's law activity students were asked to identify specific quantitative patterns using the simulation (see question 5 for this activity in the Appendix) before being asked to develop their final mathematical relationship. For Heat Capacity activity simulation, the students were also asked to identify specific quantitative patterns using the simulation but were not asked to identify specific patterns in the data provided to them (see question 4 in the Heat Capacity Activity in the Appendix) before being asked to develop their mathematical relationship. This was done is order to make the Chemistry activity maximally parallel to Physics activity and ensure that students have similar and parallel learning tasks in both activities. Since Physics activity did not have any data associated with it, adding questions on quantitative pattern identification with the data in the Chemistry activity would have made it significantly different from the Physics one.

Finally, the pre-assessment format for the two activities differed. Coulomb's law activity focused on an introductory scenario introduced prior to the simulation, and the scenario modeled the type of quantitative exploration students would eventually engage in once they started interacting with the simulation. The pre-assessment in the Heat Capacity activity, on the other hand, asked students to use the simulation right away to suggest a mathematical relationship. Considering the Energy Forms and Changes simulation was less quantitative, it is likely that the pre-assessment activity for Heat Capacity was significantly more difficult than the one for Coulomb's law. The decision to not have an introductory scenario for pre-assessment in the Heat Capacity activity was made to ensure that student have enough time to explore the simulation during the activity. The chemistry activity was implemented during a class that had ~ 50 minutes to finish the activity. Students had a 2 hour lab period to finish the physics activity.

*Implementation Context*

Both activities were implemented during Fall 2022 as part of freshmen undergraduate courses in a large public university in the Western part of the United States. The Coulomb's Law activity was implemented in an algebra-based introductory Physics course for non-Physics majors. The activity was implemented in a paper-pencil formal during the laboratory session. Students were working individually, and each turned in the assignment. A total of 73 students submitted the assignment. The Heat Capacity activity was implemented in a one-semester introductory Chemistry course for Chemistry and Biochemistry majors. The activity was implemented using Qualtrics survey tool during the regular class session. Students were working in groups of 2-3 people to complete the activity. Each group was allowed to turn in one submission. There were 29 students in the class, and they submitted 16 assignments.

*Data Analysis*

Student responses to all activity tasks were transferred to an excel spreadsheet format. Further, each student response to each MC task was scored directly into the corresponding sub-level of the cognitive framework. This was easy to do since each MC task was designed to probe specific level of the cognitive model as shown in Table 2. The pre-, mid-, and post-assessment tasks were also directly assigned a specific cognitive model level according to the rubric shown in Table 4. The rubric was previously shown to yield high inter-rater reliability (IRR) (Kaldaras & Wieman, 2023).

The main factor in level assignment was degree of sophistication of the provided justification as shown in the rubric (Table 4). Briefly, level 1 justification focuses on qualitative identification of cause-effect relationship describing the scientific phenomenon in question. Level 2 justification involves justifying the proposed mathematical formula using specific numerical values of the relevant variables. For example, at level 2 students might justify correct Coulomb's law formula by stating that this formula makes sense because you multiply given values of Q1 and Q2, divide by the squatted value of the distance between the charges, and need to multiply by Coulomb's constant. At level 2 the cause-effect relationship is still at the qualitative level, similar to that of level 1. Finally, level 3 focuses on justifying the proposed mathematical relationship by directly relating quantitative pattern observations to specific numerical operations in the equation and describing the causal mechanism described by the equation structure. For example, level 3 justification for the correctly proposed mathematical relationship for Coulomb's law would involve stating something like:

"The magnitude of electric force is linearly related to the total amount of charge on the interacting objects and inversely related to the square of the distance between the objects. This implies that the charge should be in the numerator, and the distance squared should be in the denominator of the math equation. The equation for the force should show that the force is caused by charged objects being brought close enough to interact."

Further, at the highest sub-level students should also recognize that the math equation should be multiplied by proportionality constant that ensures that the two equation sides are equal.

Table 4 shows how combinations of no, vague, or inaccurate formulas with specific types of justifications relate to specific sub-levels of the cognitive framework shown in Table 1. For example, if the accurate formula was provided, but the justification was consistent with level 1, the final level assignment was at level 1. Further, if the formula provided was inaccurate, the level assignment was also based on the degree of sophistication of the justification following the same criteria as shown in Table 4. However, in this case students would not be assigned any level beyond level 1. See the results section for examples of student responses.

The final level assignment upon completion of the activity was based on the following pieces of evidence: 1) the most sophisticated justification provided at any of the three points during the activity (pre-, -mid or -post assessment), 2) student responses to MC items probing specific sub-levels as shown in Table 2. The results section discusses student learning outcomes grounded in these pieces of evidence in more detail.

*Table 4. Scoring Rubric for Pre-, Mid- and Post- constructed response assessment tasks.*

| Accurate Formula Provided? | Elements of Student Justification | Cognitive Model Level |
|---|---|---|
| No | No justification | Level 0 (assumed the student has no idea) |
| Yes | No justification | Cannot be accurately determined |
| Yes | Justification provided at any of the sub-levels described for level 1 | The corresponding sub-level 1 |
| No | Lists only variables relevant for describing the phenomenon mathematically (Table 5 "Description" sample response) | Level 1 (Description) |
| No | Described qualitative relationships among the relevant variables (Table 5 "Pattern" sample response) | Level 1 (Patterns) |
| No | Describes qualitative causal mechanism for the phenomenon (Table 5 "Mechanism" sample response) | Level 1 (Mechanism) |
| No | Described numerical values of the relevant variables (Table 6 "Description" sample response) | Level 2 (Description) |
| No | Described quantitative patterns among the relevant variables (Table 6 "Pattern" sample response) | Level 2 (Pattern) |
| Yes | Justifies the mathematical relationship using specific numerical values of the variables (not identified) | Level 2 (Mechanism) |
| Yes | Justifies the mathematical relationship by relating the mathematical operations in the equation to specific quantitative patterns observed in the simulation and /or the data (Table 7 "Pattern" sample response) | Level 3 (Pattern) |
| Yes | Provides the causal mechanistic explanation of the equation structure (not identified) | Level 3 (Mechanism) |

**Results**

*Physics classroom: overall student progress upon completion of the activity*

The overall level assignment before and after Physics activity is shown in Figure 3.

## Physics Activity

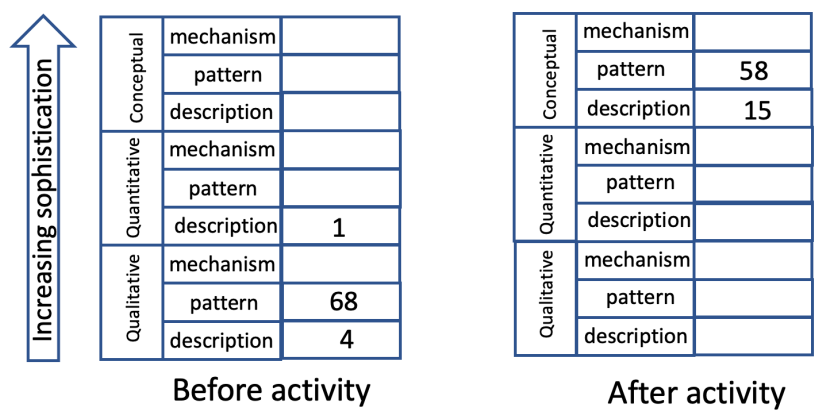

Figure 3. *Number of students and their sub-level placement before and after Physics activity.*

As shown in Figure 1, most students started at various sub-levels of level 1 (qualitative). Table 5 shows sample response for level 1 (qualitative). Most students (68) started at level 1 (qualitative) "Pattern" sub-level. An example of response at this sub-level would reflect student ability to recognize qualitative patterns among the relevant variables. Four students provided responses consistent with sub-level 1 "Description". Finally, 1 student provided the response consistent with level 2 (quantitative) "Description" sub-level shown in Table 6. Note that even if students provide any kind of mathematical relationship before the activity (vague, inaccurate, or fully accurate), the cognitive model level was assigned based on the *degree of sophistication and accuracy of their justification*. This is reflected in sample responses for each sub-level shown in Table 5. For example, sample response for level 1 (qualitative) "Mechanism" suggests an initial inaccurate formula [F= distance • q]. The justification for the formula describes which variables affect the force, which is consistent with the assigned sub-level. Level 1 (qualitative) "Mechanism" because they state a qualitative cause-effect relationship. There is no indication that the student relates the suggested mathematical relationship to observations.

Table 5. *Sample student responses for level 1 (qualitative) of the framework for Coulomb's law.*

| | | |
|---|---|---|
| **Qualitative** | Mechanism | **F=distance × q**; "the distance between the spheres and the charge affect the force" |
| | Pattern | <u>Response #1</u>: **no formula**, "as distance increases force decreases, as charge decreases, force increases"<br><u>Response #2</u>: **d/charge=f**; "as distance increases, force decreases; as charge increases, force increases" |
| | Description | **no formula**, "magnitude of force depends on distance and charge." |

Similarly, the sample response for level 2 (quantitative) "Description" also contains an inaccurate mathematical relationship in the form of [F= charge × initial separation]. The justification for this mathematical relationship states both the qualitative patterns ("when separation increases force decreases") and uses the corresponding numerical values for these variables to illustrate the numerical change ("when charge increases (1U to 9U) force increases from 0.62 N to 5.6 N"). This response is consistent with the assigned sub-level. There is no indication that the student is relating the mathematical relationship to the identified changes in numerical values of the relevant variables. Further, at level 2 (quantitative) "Pattern" sub-level students can't develop the mathematical relationship yet but demonstrate the ability to correctly identify the quantitative patterns among the relevant variables reflected in correctly answering all the parts of MC question 5 (see Appendix). Nobody demonstrated this type of blended MSS at the pre-assessment stage, but all students engaged at this sub-level during the activity and answered the MC question correctly.

Table 6. *Sample student response for level 2 (quantitative) of the framework for Coulomb's law.*

| | | |
|---|---|---|
| **Quantitative** | Pattern | Correctly answering all MC questions on quantitative pattern identification (see question 5 for this activity in the Appendix) |
| | Description | **F= charge × initial separation** Justification: "when separation increases force decreases; when charge increases (1U to 9 U) force increases (from 0.62N to 5.6 N)" |

Upon completion of the activity most students had transitioned to level 3 (conceptual) of the framework. The final level was assigned by taking into consideration the following learning indicators: 1) whether students arrived at the correct formula for Coulomb's law; 2) the level of sophistication of their justification for the formula as related to the cognitive framework levels. The highest-level justification provided at any point during the activity counted for final cognitive framework level assignment.

Importantly, all students were able to arrive at the correct formula by the end of the activity. Most of them (58) provided justification consistent with level 3 (conceptual) of the framework, which allowed to place them on a sub-level with high degree of confidence.

However, there were 15 students who provided the correct mathematical relationship, but either provided written justification consistent with level 1 (qualitative) or did not provide any justification. To gauge whether these students understood the provided mathematical relationship at a higher level, we evaluated the following indicators: 1) whether the students correctly identified quantitative relationships probed in multiple choice questions (question 5 of the Coulomb's law activity shown in the Appendix) as the indicator of level 2 (quantitative) "Pattern" sub-level and; 2) whether they recognized the need for the constant of proportionality as the indicator of level 3 (conceptual) "Description" (questions 8 and 10 of the Coulomb's law activity shown in the Appendix). The analysis showed that all 15 students were able to correctly identify the quantitative patterns, which is consistent with level 2 (quantitative) "Pattern" and recognize the need for the constant of proportionality. They also said why it is important and how to calculate it, which is the criteria of level 3 (conceptual) "Description", and hence was the level they were assigned.

By the end of the activity 58 students had transitioned to level 3 (conceptual) "Pattern" sub-level reflected in the ability to relate the observed quantitative patterns to the mathematical operations in the equation. These students presented the correct mathematical relationship, correctly identified all the relevant quantitative patterns in MC questions (Q5), recognized the need for the constant of proportionality in the equation, and demonstrated the ability to directly relate quantitative patterns to the mathematical operations in the equation in their justification. Sample response for level 3 (conceptual) "Pattern" are shown in Table 7.

Table 7. *Sample student responses for level 3 (Conceptual) for Coulomb's law.*

| | | |
|---|---|---|
| **Conceptual** | Pattern | **F= (Q1×Q2/d²) × C**<br>Justification: as distance increases, force decreases exponentially by a factor of $x^2$; as charge increases, force increases by a factor of X |
| | Description | Sample response #1:<br>**F = (Q1×Q2/d2) × C**<br>Justification: as distance increases, force decreases, as charge increases, distance increases<br>Identifying quantitative patterns identification (Q 5): all correct<br>Recognizing the need for a constant and provide sample calculations (Q 8 &10): yes<br><br>Sample response #2:<br>**F = (Q1× Q2/d2) × C**<br>Justification: none<br>Identifying quantitative patterns identification (Q 5): all correct<br>Recognizing the need for a constant and provide sample calculations (Q 8 &10): yes |

Overall, there is a clear progression of all 73 students from the lowest, level 1 to the highest, level 3 of the cognitive model. Next, we will discuss how this progress occurred during the activity focusing specifically on the tasks and points during activity at which the growth occurred. This discussion will highlight the types of learning supports within the activity that were especially useful for helping students transition to higher levels of the cognitive framework.

*Physics classroom: student learning during the activity*

In this section we will discuss the initial cognitive level assignments for students in the Physics classroom and show the learning paths that they followed during the activity. **The pie chart** in Figure 4 shows where the students in Physics class started in terms of providing the formula for Coulomb's law before doing the activity. There were three groups: 1) those that did not provide any formula; 2) those that provided vague or inaccurate formula; 3) those that provided accurate formula but no justification before the activity. Further, **the bar graphs** in Figure 4 show at what point during the activity students who provided no, vague/inaccurate, or correct formula initially attained their highest level of the cognitive model. We will further discuss each of these groups in more detail.

*Figure 4. Student learning trajectory from the beginning to the end of Coulomb's Law activity.*

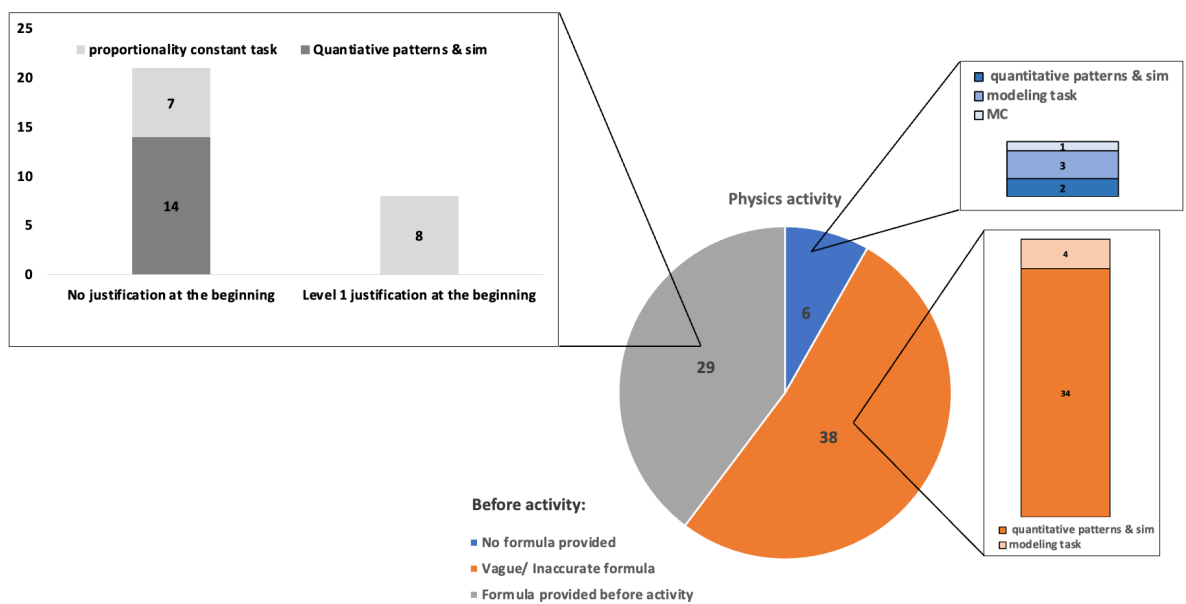

Students who did not provide a formula before the activity

There were six students who did not provide any formula before the activity (blue portion of the pie graph in Figure 4). Out of these six students:1) two students figured out the formula after interacting with the simulation and answering MC questions on quantitative patterns (Question 5); 2) three students figured out the formula after using the simulation, answering the MC questions focused on quantitative patterns and using the simulation to model the introductory scenario ( Question 5 and Task 2); 3) one student figured out the formula from a MC list of possible formulas (Question 7). All six students provided justifications consistent with level 3 (Conceptual) "Pattern". This finding suggests that the activity was helpful in supporting student progress to the highest, level 3, of the cognitive model. Moreover, the majority (5 out of 6 students) did not need MC list of possible formulas to figure out the mathematical relationship. Rather, they were able to figure out the correct formula upon completing the tasks focused on using the simulation to identify the quantitative patterns, which is reflective of engaging in blended MSS at level 2 (quantitative) "Pattern" level. Therefore, engaging in blended MSS at level 2 (quantitative) "Pattern" seems to be critical in helping students transition to level 3.

Students who provided vague or inaccurate formula before the activity

Out of 38 students who provided vague or inaccurate formula (orange portion of the pie graph in Figure 4), all figured out the formula before the MC list of possible formulas (before Question 7): 34 figured out the formula after interacting with the simulation and answering MC questions focused on noticing the quantitative patterns (Question 5). The remaining 4 students figured out the formula after using the simulation to model the introductory scenario (Question 7). All six students provided justifications consistent with level 3 (Conceptual) "Pattern". This finding suggests that the activity was helpful is supporting progress to the highest, level 3, of the cognitive model. Moreover, all 38 students needed only the tasks focused on engaging them in level 2 (quantitative) "Pattern" type of blended MSS (Questions 5 and 7) to figure out the correct formula. Nobody needed to choose out of the list of possible MC formulas to figure out the mathematical relationship.

<u>Students who provided the formula before the activity</u>

Out of 29 students who provided the formula before activity (grey portion of the pie graph in Figure 4):

***Twenty-one students provided no justification at the beginning of the activity.*** Fourteen students out of these 21 students provided justification consistent with level 3 (conceptual) "Pattern" after interacting with the simulation and answering MC questions focusing on the quantitative patterns' identification (Question 5), which is reflective of engaging in blended MSS at level 2 (quantitative) "Pattern". The remaining 7 students correctly answered the MC questions focused on identifying quantitative patterns (Question 5) and recognized the need for the proportionality constant in the equation, which is reflective of level 3 (conceptual) "Description" but did not provide any justification at any point during the activity (Questions 8 and 10). These results indicate that they can correctly identify the quantitative patterns observed in the simulation (level 2 (quantitative) "Pattern"), and they have the knowledge of the equation structure. However, there is no evidence that they can fully relate the identified quantitative patterns to the equation structure, which indicates level 3 (conceptual) "Pattern". All these pieces of evidence suggest that the highest level they demonstrate is level 3 (conceptual) "Description". *We believe that these students need further instructional supports to learn to relate the quantitative patterns to the equation structure, which would help them transition to level 3 (conceptual) "Pattern" of the cognitive model.*

 ***Eight students provided level 1 (qualitative) justification at the beginning of the activity***. All these students correctly responded to MC questions focusing on recognizing quantitative patterns and recognized the need for a constant of proportionality, suggesting the evidence for level 2 (quantitative) "Pattern" and level 3 (conceptual) "Description" respectively. However, these students didn't provide any justification relating the quantitative patterns to the structure of the mathematical equation. Therefore, these students are likely at level 3 (conceptual) "Description". *We suggest that these students need further instructional supports to help them relate the quantitative patterns to the structure of the mathematical equation, which will help them transition to level 3 (conceptual) "Pattern" of the cognitive model.* In the discussion section we will describe the kinds of instructional support that could be used in the context of this instructional approach to help these students attain the higher level.

*Chemistry classroom: overall student progress*

We followed the same process as described for Physics activity to assign cognitive framework levels to student responses on the questions probing their ability to engage in blended MSS at various levels for Heat Capacity activity. Figure 5 shows the overall student group progress by the end of the activity. Note that students in the Chemistry classroom worked in groups and submitted the activity responses in groups as well. As a result, Figure 5 shows progress of the groups of students rather than individual students.

As shown in Figure 5, most students started at various sub-levels of level 1 (qualitative). Table 8 shows sample responses for level 1 (qualitative). Additionally, 1 group provided the response consistent with level 2 (quantitative) "Pattern" sub-level shown in Table 9. By the end of the activity most groups (nine out of 15) moved to level 3 (Conceptual) Pattern as reflected in the ability to relate quantitative patterns to the equations structure (see Table 10 for sample responses). Further, 2 groups progressed to level 3 (Conceptual) "Description" as reflected in them correctly identifying the formula (see Question 6 for this activity in the Appendix) and answering the questions on quantitative patterns (see Question 4 for this activity in the Appendix), but not relating quantitative patterns to the equation structure in their justifications. These students also provided correct calculations of the specific heat capacity constants for brick, water and olive oil using the data provided to them (Questions 7 and 8) which is indicative of level 3 (Conceptual) "Description". Further, 2 groups progressed to level 2 (quantitative) "Mechanism" as indicated by them correctly identifying the formula and justifying their formula using the numerical values from the data provided to them. Finally, the remaining 2 groups correctly identified the formula by the end of the activity but did not provide justification at any point during the activity, so their post-activity level placement cannot be accurately determined. These groups also did not provide the responses to the items focusing on identifying quantitative patterns (Question 4) or calculating the specific heat capacity constants for substances (Questions 7 and 8), which makes it impossible to identify whether they have attained level 2 (quantitative) and level 3 (conceptual) respectively.

Figure 5. *Number of student groups and their sub-level placement before and after activity.*

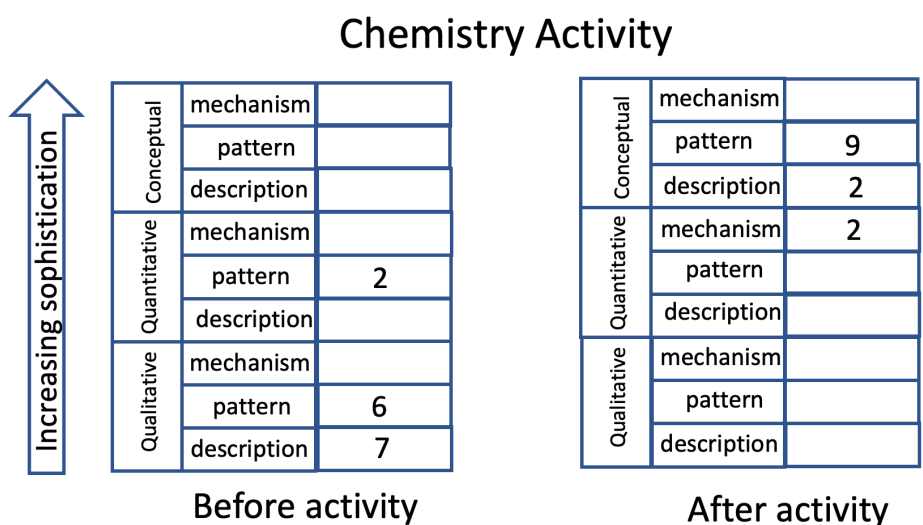

*Note: two groups did not provide justification for the formula or answers to other assessment questions which makes it impossible to determine their post-activity level*

Table 8. *Sample student responses for level 1 (qualitative) of the framework for Heat Capacity.*

| | | |
|---|---|---|
| **Qualitative** | Mechanism | An increase in given amount of a substance (g) with an increase in the change in final temperature requires a higher amount of energy to be put into the system. |
| | Pattern | As you add more mass, it will become more difficult to increase the temperature of the substance. |
| | Description | The change in temperature, mass, and volume |

Table 9. *Sample student response for level 2 (quantitative) of the framework for Heat Capacity.*

| | | |
|---|---|---|
| **Quantitative** | Mechanism | **E required to change the T of a given amount of substance = C × ΔT × m**<br>Justification: As seen the temp is dependent on the mass and not simply addition of division but multiplication of numbers. |
| | Pattern | There is a direct, linear relationship with how the temperature of a substance can be changed and with the amount of energy added to the substance |

Table 10. *Sample student responses for level 3 (conceptual) of the framework for Heat Capacity.*

| Conceptual | Pattern | E required to change the T of a given amount of substance = C × ΔT × m<br><u>Justification</u>: Temperature change by 10 changed the joules in the system by a factor of 10. If you double the mass of the substance, it doubles the joules in the system. |
|---|---|---|
| | Description | provided correct calculations of the specific heat capacity constants for brick, water and olive oil using the data provided |

Overall, there is a clear progression of all the groups to the higher levels of the cognitive framework, except for the two groups that did not provide all responses. Most groups (11 out of 15) progressed from the lowest, level 1, to the highest, level 3. Next, we will discuss how this progress occurred during the activity focusing specifically on the tasks that promoted the growth.

*Chemistry classroom: student learning during the activity*

The group learning trajectories in the Chemistry classroom are shown in Figure 6. In Chemistry activity only 1 group started with providing the correct formula, but they didn't provide any justification for it. This group later provided justification at level 2 (quantitative) "Mechanism" level by grounding their suggested formula in the numerical values of the data provided to them (see Table 9 "Mechanism" sample response). The justification was provided at the end of the activity after having explored the simulation, the data, and the list of possible MC formulas. The remaining 14 groups started with no formula initially. Out of 14 groups, 5 groups developed the correct mathematical relationship upon completing the tasks on identifying the quantitative patterns (Question 4) and exploring the data provided to them. These five groups therefore didn't need to see the list of possible formulas to develop the correct mathematical relationship. The remaining nine groups identified the correct mathematical relationship from the MC list of possible formulas provided to them (Question 6). This is the main difference between the Chemistry and Physics classrooms: more students in the Chemistry classroom needed to see the list of possible formulas to identify the correct mathematical relationship as compared to Physics classroom. In contrast, in the Physics classroom most students identified the mathematical relationship by exploring the simulation prior to seeing the list of possible formulas. We discuss the possible reasons and implications of this finding in the next section.

Figure 6. *Student learning trajectory from the beginning to the end of Heat Capacity activity*

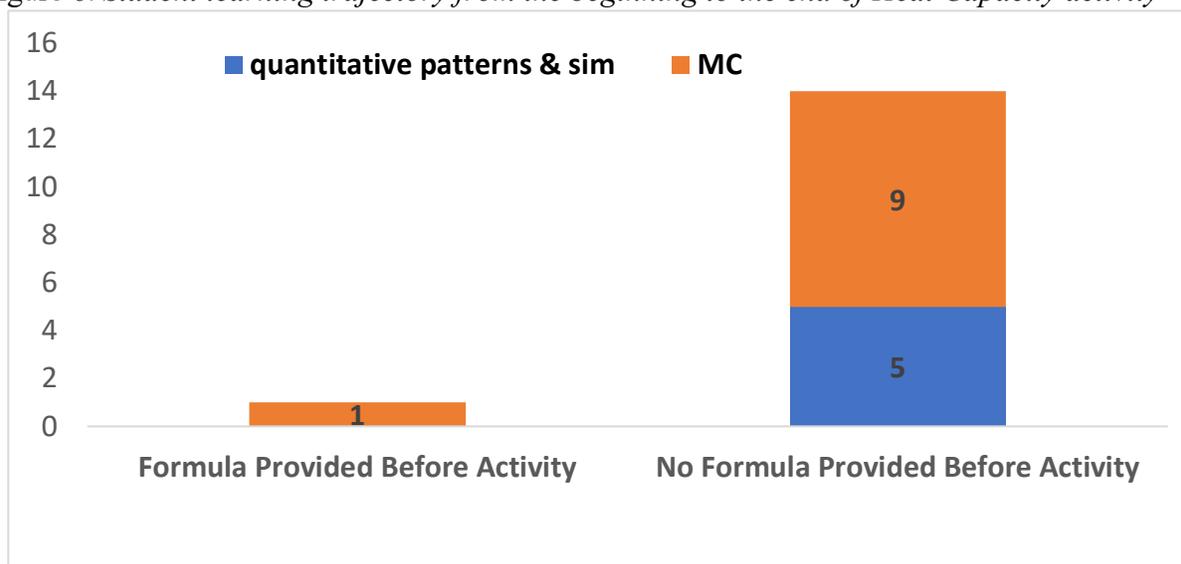

**Discussion**

We have presented an instructional model for teaching blended MSS in introductory undergraduate STEM courses. The instructional model is grounded in previously validated cognitive framework for blended MSS (Table 1). The instructional model presented here can be used to design learning sequences supporting student in developing blended MSS skills across STEM disciplines at the undergraduate level. We also believe that this model is applicable to K-12 settings and fully aligned with the principles of authentic science learning outlined in the Framework for K-12 Science Education (National Research Council, 2012). However, future work is needed to demonstrate the effectiveness of this model at the K-12 settings.

We have demonstrated that the learning sequences developed following this instructional model for Coulomb's Law in Physics and Heat Capacity in Chemistry help most undergraduate students transition from low to the highest levels of the cognitive framework. Moreover, we discovered that students in both classes started mostly at the lowest level of the cognitive framework, level 1. This indicates that students had very little understanding of the mathematical relationship underlying the phenomenon under study and how the relationship described the scientific observations but made great progress in completing the learning sequence. Student response evidence indicates that most students were able to successfully engage in blended MSS at the highest level (conceptual) "Pattern" type of sensemaking. Therefore, these results indicate that the instructional sequences developed following the instructional model grounded in the validated cognitive framework for blended MSS (Kaldaras & Wieman, 2023) are effective in helping student develop higher proficiency in blended MSS, which addresses RQ1 of the study.

Aligning the learning tasks with specific sub-levels of the cognitive framework ensures that instructors can identify specific ideas that students are struggling with (Kaldaras et al., 2021). The current study demonstrated that one of the most common student struggles in both classes was relating the quantitative patterns identified using the simulation to the equation structure, which is reflective of level 3 (conceptual) "Pattern" type of blended MSS. While most students in both classes were able to attain this sub-level by the end of the activity, there were still a significant number of students who didn't provide justifications for their formula consistent with this sub-level. However, those students still demonstrated engagement in blended MSS at all the lower sub-levels, indicating considerable progress. This demonstrates that the instructional approach grounded in the cognitive framework is effective for both organizing and guiding instruction: it offers actionable information on student proficiency in blended MSS to guide how to support students to overcome their difficulties. For example, an instructor might choose to spend additional time on helping students transition to level 3 (conceptual) "Pattern" by providing additional learning opportunities focused on helping student make connections between the quantitative patterns consistent with level 2 (quantitative) "Pattern" and the equation structure for various phenomena.

Engaging students in level 2 (quantitative) "Pattern" type of blended MSS seems to be critical to helping them transition to level 3. Notice that once students have completed tasks focused on quantitative pattern identification aimed at engaging them in level 2 (quantitative) "Pattern" type of blended MSS, most students in both activities proceeded to developing the mathematical relationship as shown in Figures 4 and 6 for Physics and Chemistry activity respectively. This finding indicates that learning tasks focused on engaging students in identifying specific quantitative patterns are effective in helping them reach level 3 and build connection between the quantitative patterns and the mathematical structure of the underlying relationship. This answers RQ2 of the study *RQ2: What instructional elements are helpful in supporting students' progress along the levels of the cognitive model?*

Another important aspect of this instructional approach is grounding the blended MSS in student exploration of the interactive computer simulations, such as PhET simulations. Evidence from the current study suggests that PhET simulations are effective in supporting student engagement in blended MSS at all the sub-level of the cognitive framework shown in Table 1. Specifically, PhET sims offer an authentic way to explore the phenomenon in question, identify the relevant variables and their associated numerical values (level 1 and 2 "Description" respectively), gauge the scientific and mathematical meaning of the unobserved variables (level 3 "Description"), explore qualitative and quantitative relationships among the relevant variables (level 1 and 2 "Pattern" respectively), investigate the cause and effect relationships among the relevant variables to develop qualitative or quantitative understanding of the causal mechanism (level 1 and 2 "Mechanism" respectively), build a deeper understanding of the relationship between the quantitative patterns and the mathematical operations in the formula (level 3 "Pattern") and the causal structure of the mathematical relationship (level 3 "Mechanism"). These unique feature of PhET simulations remove cognitive complexities of static learning environments, such as the need to verbally describe the phenomenon, introduce data tables reflecting relevant quantitative patterns among the variables, or encounter measurement errors associated with hands-on experiments, all of which can significantly complicate engagement in blended MSS.

Moreover, the data suggests that the simulations that allow for direct and obvious exploration of the quantitative patterns among the variables might be more effective in helping students transition to level 3 than those that don't. This is reflected in the difference between the learning trajectories among the Physics and Chemistry students: most Physics students developed the mathematical relationship right after tasks focused on identifying quantitative patterns among the relevant variables (Figure 4), while most of the Chemistry students needed to see the MC list of possible formulas to finalize the mathematical relationship in addition to completing tasks on quantitative pattern identification (Figure 6). Even though the chemistry simulation was supplemented with additional data tables showing the amount of energy it takes to heat up different amounts of the four substances by various number of degrees most student groups (10 out of 15) did not develop the formula and justification from exploring the data and the simulation. This finding suggests that a simulation that does not support straightforward exploration of quantitative patterns (level 2 (quantitative) "Pattern" type of sensemaking), even with supplementary data, is not as effective in helping student develop the mathematical relationship and progress to level 3.

An alternative explanation for the observed differences is that students in Chemistry classroom were not able to engage in effective quantitative pattern exploration because they were not explicitly asked to identify quantitative patterns among the relevant variables in the provided data but were only asked to do that in the simulation (Question 4), which does not allow for a straightforward identification of the patterns. Therefore, future research should focus on introducing slight modifications to the Chemistry activity focused on adding MC questions asking students to identify specific quantitative patterns in the data, parallel to Question 5 in the Physics activity. Introducing this instructional support could potentially be sufficient for students to develop the mathematical relationship upon exploring the data and without the need to see the MC list of possible formulas.

While most students in both classes developed the mathematical relationship and provided justification consistent with level 3 (conceptual) "Pattern" by the end of the activity, there were still a number of students who either provided no justification or provided justification consistent with level 1 (qualitative) by the end of the activity. As mentioned above, these students need additional learning supports to build understanding of the mathematical relationship in relation to the scientific phenomenon. These learning supports should focus on helping students build connections between the quantitative patterns and the equation structure, including all the relevant variables and the mathematical operations among the variables.

Finally, the data show that no student attained the highest sub-level of the framework-level 3 (conceptual) "Mechanism". This finding suggests that simply asking students to justify their proposed relationship and explain the causal mechanism described by the relationship is not sufficient for eliciting this level of blended MSS. A way to address this shortcoming might be to scaffold student exploration of the simulation and data (when applicable) to help students identify specific causal relationships and relate them back to the equation structure.

There are several limitations of this work that are important to mention. First, the instructional approach presented in this study is demonstrated using a small number of topics and students. In the future it would be beneficial to expand the design of the learning sequences to include more relevant topics across various STEM disciplines, including Chemistry, Physics and Biology among others and pilot the sequences with larger number of students. Additionally, future research needs to address ways of helping students explicitly engage in level 3 (Conceptual) type of blended MSS as they pursue the instructional sequences. More research is needed on designing and evaluating the effectiveness of instructional supports for engaging students in blended MSS at the highest sub-level of the framework across STEM topics and disciplines. Finally, the learning approach needs to be extended to support engagement in blended MSS with other important mathematical tools, such as vectors and vector operations.

To summarize, this work introduces a first-ever published instructional approach for supporting the development of blended MSS skills across STEM disciplines. The instructional approach is grounded in previously validated cognitive framework and as demonstrated in this study is a straightforward and effective way of helping students develop higher proficiency in blended MSS in various STEM disciplines, including Physics and Chemistry.

## Acknowledgements

We would like to thank the University of Colorado Boulder Physics and Chemistry instructors, Dr. Eleanor Hodby and Dr. Robert Parson for providing feedback on the design of the instructional sequences and piloting the sequences in their classrooms. We would also like to thank Dr. Georg Rieger from the University of British Columbia for providing feedback on the design of the learning sequences and sharing his pedagogical and subject-matter expertise at all stages of this project. This work would not be possible without their support, input and encouragement. We would also like to thank the PhET Simulations Project team for making this wonderful educational tool freely available around the world. This work was supported by the Yidan Foundation.

**Appendix**

<span style="color:red">Text in red described the alignment between the questions and the levels of the cognitive framework shown in Table 1 as well as the activity structure shown in Table 2 in the main document.</span>

**Physics Activity**
## What quantities does electric force depend on?

Consider the following experiments performed on 2 charged metal spheres floating in space (ignore gravity) like in the Phet simulation (introduced below):

- Case 1. Both metal spheres have 1 unit of charge (U). When placed 4.0 cm apart the electric force experienced by each of them is measured to be of magnitude 5.6 N.
- Case 2. When the initial separation is increased to 8.0 cm, the electric force experienced by each charge becomes about 1.4 N.
- Case 3. When the initial separation is increased to 12.0 cm, the electric force experienced by each charge becomes 0.62 N.
- Case 4. When the initial separation is 8.0 cm and the charge is increased from 1U to 4U on one of the spheres, the electric force experienced by each sphere is measured to have magnitude 5.6 N.
- Case 5. When the initial separation is 12.0 cm and the charge is increased from 1U to 9U on one of the spheres, the electric force experienced by each sphere is measured to be 5.6 N.

These experiments and the observed results are summarized in the picture below:

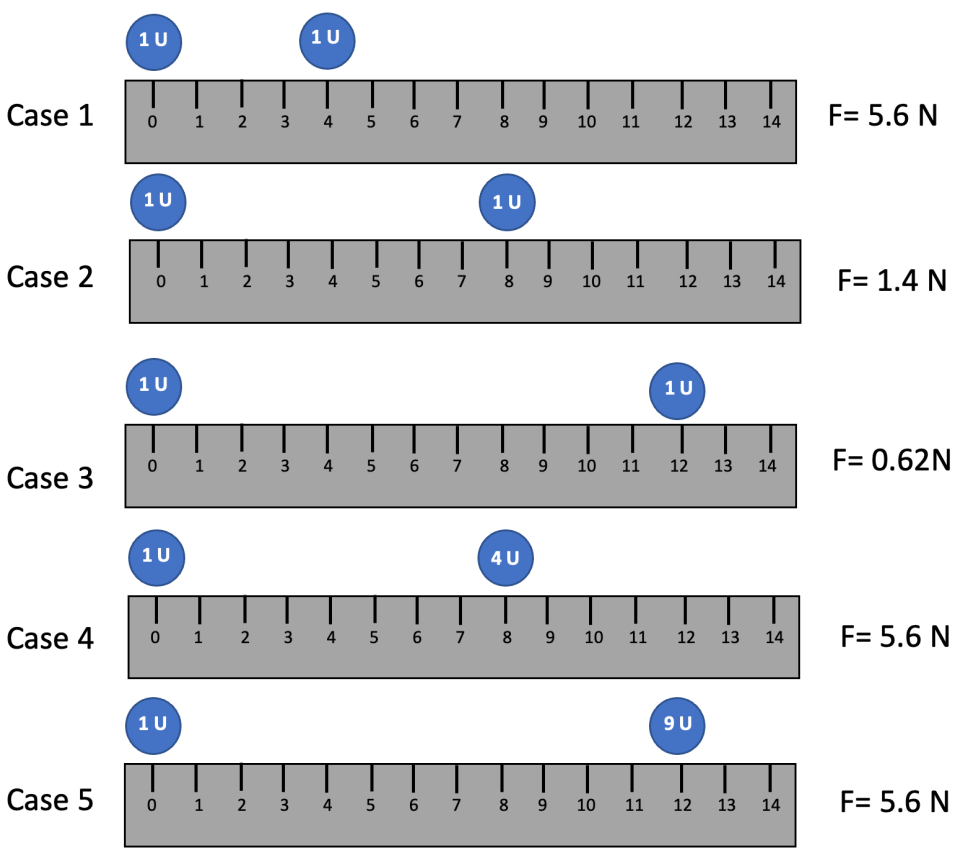

**Pre-Assessment (to gauge the starting before the activity)**

**Question 1**. Based on this data, what quantities does the magnitude of the electric force experienced by both spheres seem to depend on? Use the given data to justify your answer.

**Question 2**. Try to describe this dependence in mathematical terms (i.e. using a mathematical formula or equation) using the given data to justify your answer. Don't worry if this is really challenging at this point, just do your best and move on.

Now explore the simulation (Coulomb's law, macro-scale):

https://phet.colorado.edu/sims/html/coulombs-law/latest/coulombs-law_en.html

***Your overarching task in this activity is to develop a mathematical relationship that describes the physical phenomenon you are observing in the simulation. Start by responding to the questions below.***

**Level 1 (qualitative) "Description"**

**Question 3**. Circle or highlight the variables required to describe the phenomenon in PhET simulation mathematically:
- A. Charge on sphere 1
- B. Charge on sphere 2
- C. Distance between the spheres
- D. Electric Force
- E. Mass of the characters pushing the spheres
- F. Mass of sphere 1
- G. Mass of sphere 2
- H. Other

If other, please explain:

**Level 1 (qualitative) "Pattern"**

**Question 4**. Circle or highlight the words in each statement that describe your observations of the sim:
- A. As the distance between the two charges increases, the magnitude of the force between them [increases/decreases/stays the same]
- B. As the magnitude of the two charges increases, the magnitude of the force between them [increases/decreases/stays the same]
- C. The sign of the charge on each sphere [has/does not have] an effect on the direction of the resulting force between the charges
- D. The sign of the charge on each sphere [has/does not have] an effect on the magnitude of the resulting force between the charges
- E. The magnitude of charge on each sphere [has/does not have] an effect on the direction of the resulting force between the charges
- F. The magnitude of charge on each sphere [has/does not have] an effect on the magnitude of the resulting force between the charges
- G. If Q1> Q2, then the electric force experienced by Q1 will be [ larger/smaller/the same] as the electric force experienced by Q2.

**Level 2 (quantitative) "Pattern"**

**Question 5**. Use the simulation to collect data to answer the following questions. Circle or highlight one word in each bracket, so that the sentence describes your observations of the sim.
- A. As the distance between the two charges [increases/decreases] by a factor of [two/three/four/five/six], the magnitude of the force between them [increases/decreases/stays the same] by a factor of [one/two/three/four/five/six].

B. As the distance between the two charges [increases/decreases] by a factor of X, the magnitude of the force between them [increases/decreases/stays the same] by a factor of [ X / 2X / 3X / 4X/ 5X / 6X / X^2 / X^3 ]

C. As the magnitude of Q1 [increases/decreases] by a factor of  [two/three/four/five/six], the magnitude of the force between them [increases/decreases] by [two/three/four/five/six].

D. As the magnitude of Q2 [increases/decreases] by a factor of  [two/three/four/five/six], the magnitude of the force between them [increases/decreases] by [two/three/four/five/six].

E. As the magnitude of Q1 [increases/decreases] by a factor of X, the magnitude of the force between them [increases/decreases/stays the same] by a factor of [ X / 2X / 3X / 4X/ 5X / 6X / X^2 / X^3 ]

F. As the magnitude of Q2 [increases/decreases] by a factor of X, the magnitude of the force between them [increases/decreases/stays the same] by a factor of [ X / 2X / 3X / 4X/ 5X / 6X / X^2 / X^3 ]

## Mid-Assessment (Assess the level upon exploring quantitative patterns)
**Question 6.** Using the simulation and the patterns you have identified on the previous page, what mathematical relationship would describe the phenomenon you are observing in the simulation? Justify your answer.

## Mid-Assessment (Assess the level upon exploring quantitative patterns)
***Task 2: Use the simulation to model the 6 different cases on page 1. Try to write down an equation that would describe the dependence of the electric force on the relevant variables. Use the data from the simulation to support your proposed mathematical relationship.***

## Post-Assessment (Assess the level upon exploring quantitative patterns)
**Question 7**. Which mathematical relationship most likely explains your observations in the sim? Justify your response.

*Note: symbol "$\propto$" stands for proportionality*

$$a)\ Force \propto \frac{distance \times 2}{Q1 \times Q2} \qquad\qquad f)\ Force \propto \frac{(Q1-Q2)}{distance^3}$$

$$b)\ Force \propto \left(\frac{distance}{2}\right) \times Q1 \times Q2 \qquad g)\ Force \propto \frac{Q1 \times Q2}{distance \times 6}$$

$$c)\ Force \propto \left(\frac{1}{distance}\right) \times Q1 \times Q2 \qquad h)\ Force \propto \frac{Q1 \times Q2}{distance^2}$$

$$d)\ Force \propto \left(\frac{1}{distance^2}\right) \times (Q1 + Q2) \qquad i)\ Force \propto \frac{(Q1+Q2)}{distance^4}$$

$$e)\ Force \propto \frac{Q1 \times Q2}{distance^4}$$

## Level 3 (conceptual) "Description"
**Question 8.** How can you transition from the most plausible proportional relationship you have determined above to a complete mathematical equation? What information are you missing? Continue to explore the simulation to help you answer this question.

**Post-Assessment (Assess the level upon exploring quantitative patterns)**

**Question 9**. Which formula most likely explains your observations in the sim?
*Note: C is a constant to be determined from measurement.*

$$a) \; Force = \frac{distance \times 2}{Q1 \times Q2} \times C \qquad\qquad f) \; Force = \frac{(Q1 - Q2)}{distance^3} \times C$$

$$b) \; Force = \left(\frac{distance}{2}\right) \times Q1 \times Q2 \times C \qquad g) \; Force = \frac{Q1 \times Q2}{distance \times 6} \times C$$

$$c) \; Force = \left(\frac{1}{distance}\right) \times Q1 \times Q2 \times C \qquad h) \; Force = \frac{Q1 \times Q2}{distance^2} \times C$$

$$d) \; Force = \left(\frac{1}{distance^2}\right) \times (Q1 + Q2) \times C \qquad i) \; Force = \frac{(Q1+Q2)}{distance^4} \times C$$

$$e) \; Force = \frac{Q1 \times Q2}{distance^4} \times C$$

Justify your response.

**Level 3 (conceptual) "Description"**

**Question 10**. Based on your interactions with the simulation, what is the value and units of the constant C? Explain how your value was determined using the simulation.

**Post-Assessment (Assess the level upon activity completion)**

**Question 11**. Summarize your work so far:
   A.  Explain in words which quantities the electric force between two charged spheres depends on
   B.  Describe this relationship using math (as an equation) instead of words.

**Chemistry Activity**
**What quantities affect how a substance changes temperature?**

You know from everyday life that substances are affected by heat in different ways: a metal teapot heats up before the water in it starts to boil, and it takes less time to fry any food in oil than to boil it in water. So, what quantities affect how easily the temperature of a substance can be changed and how can the process of raising the temperature of a substance by a certain amount be described mathematically? To investigate these questions, start by exploring the following simulation.

Energy Forms and Changes: Intro
*Note: Assume the mass of brick and iron cubes is 1 kg and the mass of water and oil is also 1 kg.*

**Pre-Assessment (to gauge the starting before the activity)**
**Question 1.** Use your observations in the simulation to predict a mathematical relationship that would explain how easily the temperature of a substance can be changed. Justify your prediction.

**Level 1 (qualitative) "Description"**
**Question 2**. Select the variables that affect how much heat needs to be supplied to a substance in order to change its temperature:
  A. Type of substance
  B. Starting temperature
  C. Final temperature
  D. Total change in temperature
  E. Mass of the substance
  F. Volume of the substance
  G. Whether the substance is cooling down or heating up
  H. Heat energy

**Level 1 (qualitative) "Pattern"**
**Question 3**. Use the simulation to complete the following questions.
  A. It takes the most amount energy to heat 1 kg of (water/oil/brick/iron) by 1 degree
  B. It takes the least amount of energy to heat 1 kg of (water/oil/brick/iron) by 1 degree
  C. It takes more energy to heat up 1 kg of (water/oil) by 1 degree as compared to (water/oil)
  D. It takes more energy to heat up 1 kg of (brick/iron) by 1 degree as compared to (brick/iron)
  E. The maximum temperature of liquid water is (smaller/larger/same) as the maximum temperature of liquid oil
  F. It takes the same amount of heat to raise the temperature of any substance by one degree (True/False)
  G. The amount of heat it takes to raise the temperature of a substance by 1 degree depends on the initial and final temperature of the substance (True/False)

**Level 2 (quantitative) "Pattern"**
**Question 4**. Use the simulation to complete the following questions.
  A. It takes (2/3/4) times (more/less) energy units to heat 1 kg of water by 1 degree as compared to oil.
  B. It takes (2/3/4) times (more/less) energy units to heat 1 kg of brick by 1 degree as compared to iron.

Explore the following data.

Table 1. Water

| Mass (g) | Initial Temperature (K) | Final Temperature (K) | Energy Used (joules) |
|---|---|---|---|
| 1000 | 274 | 275 | 4180 |
| 1000 | 274 | 284 | 41800 |
| 2000 | 274 | 275 | 8360 |
| 2000 | 274 | 284 | 83600 |

Table 2. Oil

| Mass (g) | Initial Temperature (K) | Final Temperature (K) | Energy Used (joules) |
|---|---|---|---|
| 1000 | 274 | 275 | 2000 |
| 1000 | 274 | 284 | 20 000 |
| 2000 | 274 | 275 | 4000 |
| 2000 | 274 | 284 | 80 000 |

Table 2. Brick

| Mass (g) | Initial Temperature (K) | Final Temperature (K) | Energy Used (joules) |
|---|---|---|---|
| 1000 | 274 | 275 | 840 |
| 1000 | 274 | 284 | 8400 |
| 2000 | 274 | 275 | 1680 |
| 2000 | 274 | 284 | 16800 |

Table 3. Iron

| Mass (g) | Initial Temperature (K) | Final Temperature (K) | Energy Used (joules) |
|---|---|---|---|
| 1000 | 274 | 275 | 412 |
| 1000 | 274 | 284 | 4120 |
| 2000 | 274 | 275 | 824 |
| 2000 | 274 | 284 | 8240 |

**Mid-Assessment (Assess the level upon exploring quantitative patterns using the sim and the data)**

**Question 5**. Use your observations in the simulation and the data to predict a mathematical relationship that would explain how easily the temperature of a substance can be changed. Justify your prediction.

**Post-Assessment (Assess the level upon activity completion)**

**Question 6**. Select a proportional relationship that most likely explains how easily the temperature of a substance can be changed. Justify your choice.

    A.  E required to change the T of  a given amount of substance = C × **Δ**T
    B.  E required to change the T of a given amount of  substance = C × **Δ**T × m
    C.  E required to change the T of a given amount of  substance = C ÷ **Δ**T
    D.  E required to change the T of a given amount of  substance = C ÷ m
    E.  E required to change the T of a given amount of  substance= C ÷ (**Δ**T ÷ m)
    F.  E required to change the T of a given amount of  substance = C+ **Δ**T+ m
    G.  E required to change the T of a given amount of  substance = C+ **Δ**T
    H.  E required to change the T of a given amount of  substance = C - **Δ**T
    I.  E required to change the T of a given amount of  substance = C - **Δ**T+ m

Note: E= energy
       C= constant to be determined from measurement
       **Δ** = change in (delta)
       T= temperature
       m=mass

**Level 3 (conceptual) "Description"**

**Question 7**. Based on the data provided in tables 1-4, calculate the value of the constant for each substance (water, oil, brick, iron). Show your work.

**Level 3 (conceptual) "Description"**

**Question 8**. Explain the scientific meaning of the constant you have calculated for each substance.